       \documentstyle[astrobib,epsf]{mn}
\begin{document}

\def\lesssim{\, \lower2truept\hbox{${<\atop\hbox{\raise4truept\hbox{$\sim$}}}$}\,}
\def\gtrsim{\,\lower2truept\hbox{${> \atop\hbox{\raise4truept\hbox{$\sim$}}}$}\,}
\def\micron{\mbox{$\mu$m}}

\def\fdie{F10214+4724}
\def\inov{09104+4109}
\def\fqui{F15307+3252}
\def\mmu{$m^{-1}$}
\def\mmum{$m^{-1/2}$}
\def\hmd{$h^{-2}$}
\def\hmu{$h^{-1}$}

\title[Supermassive Black Holes In Early Type Galaxies]
{Supermassive Black Holes in Early-Type Galaxies: Relationship with Radio 
Emission and Constraints on the Black Hole Mass Function}

\author[A. Franceschini, S. Vercellone, A. Fabian]
{A.\ Franceschini$^1$, S.\ Vercellone$^1$ and A.C. Fabian$^2$ \\ 
$^1$Dipartimento di Astronomia, Vicolo dell'Osservatorio 5, I-35122
Padova, Italy\\
Tel. (049) 8293441, Fax (049) 8759840,
E--mail "Franceschini@pdmida.pd.astro.it"\\
$^2$Institute of Astronomy, Madingley Road, Cambridge CB3 0HA \\
Address for correspondence: A. Franceschini}

\maketitle

\begin{abstract}
Using recently published estimates - based on high spatial resolution 
spectroscopy - of the mass $M_{\rm BH}$ of nuclear black
holes for a sample of nearby galaxies, 
we explore the dependence of galaxy nucleus emissivity at
various wavelengths on $M_{\rm BH}$. We confirm an almost linear scaling of
the black hole mass with the baryonic mass of the host spheroidal galaxy.
A remarkably tight relationship is also found with both nuclear and total
radio centimetric flux, with a very steep dependence of the radio flux on
$M_{BH}$ ($P\propto M_{\rm BH}^{2.5}$).  The high-frequency radio power
is thus a very good tracer of a super-massive black hole, and a good
estimator of its mass. This, together with the lack of significant
correlations with the low-energy X-ray and far-IR flux, supports the view
that advection-dominated accretion is ruling the energy output in the
low-accretion rate regime.  Using the tight dependence of total radio
power on $M_{\rm BH}$ and the rich statistics of radio emission of
galaxies, we derive an estimate of the mass function of remnants in the
nearby universe.  This is compared with current models of quasar
and AGN activity and of the origin of the hard X-ray background (HXRB).
As for the former, continuous long-lived AGN activity is excluded by the
present data with high significance, whereas the assumption of a 
short-lived, possibly recurrent, activity pattern gives remarkable
agreement. The presently estimated black hole mass function also implies that
the HXRB has been produced by a numerous population ($\sim 10^{-2}\ Mpc^{-3}$)
of moderately massive ($M_{\rm BH}\sim 10^7\ M_\odot$) black holes.

\end{abstract} 


\section{Introduction}

The existence of a massive black-hole (BH) in galaxy nuclei (see e.g.
Rees 1984) is the
current theoretical paradigm for interpreting a variety of astrophysical
phenomena, including activity in optical and radio quasars, Seyfert galaxy
nuclei, and in various classes of radio galaxies, blazars, and
hyperluminous IRAS galaxies.

Powerful spectrographs and high-resolution imagers on optical (HST) and
X-ray telescopes (ROSAT, ASCA) now allow the existence of such massive
black holes ($M_{BH} \sim 10^6 - 3\ 10^9\ M_\odot$) to be tested in the
nuclei of a few nearby galaxies, via gas and stellar dynamics, including
the evidence for relativistic motions (Tanaka et al., 1995; see a review
in Kormendy \& Richstone, 1995). The results are consistent with the
general BH interpretation of nuclear activity.

However, several details in this picture are still missing. One is the
marked dichotomy in objects presumed to harbour a nuclear BH: at large
distances and high redshifts we find strong evidence of activity
associated with massive BHs, but no direct proof of them, mostly because the
search for a BH requires resolving the radius of influence of the BH,
and then is possible only out to few tens of Mpc. 
On the contrary, in local objects good dynamical evidence for the
presence of nuclear BHs is found, but it is mostly unrelated to the
monster AGN activity seen at high z.

Other missing details include the formation and evolution of supermassive
BH and their relationship with the host galaxy population in general.
Specific predictions of physical models of quasars and quasar evolution
(e.g. Soltan, 1982; Chokshi \& Turner, 1992; Cavaliere \& Padovani,
1989) suggest that the AGN
phenomenon is short-lived and probably recurrent, with the implication
that a significant fraction of all massive galaxies should contain a BH of
typically $\sim 10^8-10^9 M_\odot$, which is currently inactive.

Further arguments in favour of a deep relationship between quasar
activity and the formation of a massive spheroidal galaxy have been given
by Hamann \& Ferland (1993) and Franceschini \& Gratton (1997, FG97),
based on the chemical properties of the enriched gas in the vicinity of
the QSO. A one-to-one relationship of quasars with forming spheroids is
suggested by FG97, from the consideration that not only all
observed QSOs show large amounts of stellar-processed material in the 
nucleus, but also that the
significantly metal-enriched sites at high redshifts are only found close
to a QSO. The presence of a quasar may hide an underlying spheroid in the
process of formation, thus possibly explaining the long-standing problem
arising from the lack of evidence for galaxies forming at high $z$ (e.g.
Djorgowski, 1995; FG97; see also, in a different context, Terlevich
\& Boyle, 1993).

All these arguments naturally imply that most spheroidal galaxies today
should harbour a super-massive BH in their nuclei. This raises a 
potential problem for the BH interpretation: under this scheme, a
lot more X-ray emission by Bondi-accreted halo plasma would be
expected in elliptical galaxies than is observed (Fabian \& Canizares,
1988).

This is now explained (Fabian \& Rees 1995; Di Matteo \& Fabian 1996a) in
terms of an advection-dominated accretion flow (ADAF: Rees, 1982; Begelman
1986; Narayan \& Yi 1995; Narayan 1995,1996) occurring at low accretion
regimes $\dot m$, in place of the traditional thin-disc accretion mode. In
such a case the radiative efficiency of the accreting material is low, and
the energy released by viscous friction is advected into the BH rather than
radiated.  Massive BHs would then be very dim at the frequencies (IR,
optical, soft X) at which typical AGN with higher accretion rates are very
prominent.

Using recently published estimates of nuclear BH masses in nearby
galaxies, we explore in this paper the efficiency of galaxy emission
properties at various wavelengths (radio, X-ray, optical and IR) to trace
the presence of supermassive BHs in the local universe. Correlations
between the estimated BH masses and electromagnetic emissions at various
wavelengths are also used to constrain the physical conditions of the
accreting gas, with specific reference to the ADAF model. Finally we
derive the BH mass function and compare it with theoretical distributions
based on quasar statistics and related arguments.

In Section {\bf II} we summarize the optical, X-ray and radio information
available for each galaxy in our sample. In Section {\bf III} we discuss
the relationship between the BH mass and e.m. emission, with an emphasis
on the radio power. Section {\bf IV} addresses the question of
advection-dominated accretion models, which appear to receive some
support by the present analysis. In Section {\bf V} we estimate the BH
mass function derived from the previous regressions and the galaxy local
luminosity functions in the optical and the radio. In Section {\bf VI} we
discuss our results.

A value of $H_0=50$ km~s$^{-1}$~Mpc$^{-1}$ is adopted throughout the
paper.

\section{THE DATA}

Our reference sample for BH mass estimates in nearby galaxies is based on
the 8 galaxies reported by Kormendy \& Richstone (1995, KR95). 
We have improved
it by adding five new objects with recently published data, as reported
in the lower part of Table~1. We have adopted a quality flag (QF) for the
BH mass evaluation (column 12 in Table 1). QF=3 corresponds to the safest
case of a mass determined with high spatial resolution spectroscopy.
Lower values for QF correspond to decreasing reliability.

In the following we review observational data collected from the literature
for each galaxy. 

\subsection{Galaxies from the KR95 compilation}

{\bf M31. }{\it Radio} data come from Becker \& White (1991; BW91 hereafter) 
and refer to the total emission from the galaxy; {\it IR} data are from 
Rice et al. (1988), while {\it X-ray} data from Brinkmann et al. (1994). 
The value of BH mass comes from KR95, based on the M/L ratio and from the fact
that M31 seems to harbour a double nucleus.

{\bf NGC 3115. }{\it Radio} data come from KR95 and refer to the nuclear radio
continuum flux; {\it IR} data are from van
Driel et al. (1993) and {\it X-ray} data from Fabbiano et al. (1992). The most
recent evaluation of the BH mass comes from Kormendy et al. (1996a) 
based on  HST-FOS nuclear spectroscopy and on modelling
of the stellar velocity and velocity dispersion curves.

{\bf M32. }{\it Radio} flux comes from imaging photometry at 4.850 GHz 
by Condon et al. (1994) and refers to the total emission; {\it IR} flux is from
Knapp et al.. (1989). The BH mass is from Bender et al. (1996), 
based on CFHT high spatial resolution spectroscopy and van der Marel et al. 
(1997) from HST data. 

{\bf NGC 4594.} {\it Radio} data are from Griffith et al. (1994); {\it
IR} data from Condon et al. (1995), {\it X-ray} data from Fabbiano et al.
(1992). An updated value of the BH mass comes from modelling the
stellar velocity distribution as measured with HST and CHFT (Kormendy et
al., 1996; Fabbiano \& Juda, 1997). 

{\bf Milky Way. }{\it Radio} luminosity is reported by KR95 and refers to 
the nuclear continuum. The value of BH mass comes from Eckart \& Genzel
(1997) based on proper motions of stars in the innermost core of the Galaxy.

{\bf NGC 3377. }{\it Radio} data are from Dressler \& Condon (1978, DC78) 
and refer to the total emission; {\it IR} flux is from 
Peletier et al. (1990), the {\it X-ray} from Fabbiano et al. (1992). The 
value of BH mass comes from modelling by KR95 of the stellar velocity 
dispersion.

{\bf NGC 4258. } The total {\it radio} flux is from Gregory \& Condon
(1991) (GC91 hereafter) the {\it IR} flux is from Ghosh et al. (1993);
the {\it X-ray} flux from Fabbiano et al. (1992). The value of BH mass is
given by Miyoshi et al (1995) fitting the Keplerian motion of water
masers. This value has been confirmed by Maoz (1995) and Wilkes et al.
(1995). The latter report the optical detection of the hidden nuclear
engine. Lasota et al. (1996) have shown that the observed spectrum is
well fitted by an advection-dominated disc emission with BH mass
consistent with our adopted value.

{\bf M87. }{\it Radio} data come from GC91 and refer to the total
emission from the galaxy; {\it IR} data are from Condon et al. (1994) and
the {\it X-ray} from Brinkmann et al. (1994) (the latter however is
mostly contributed by the intracluster gas). The value of the BH mass
comes from Ford et al. (1994) based on gas-dynamical evidence of a
nuclear disc of ionized gas in Keplerian motion. The mass value has been
found by Reynolds et al. (1996) to be consistent with the theoretical
expectation based on an ADAF model.

\subsection{Further objects with nuclear BH mass estimates }

{\bf NGC 4261. }{\it Radio} data are published by Griffith et al. (1995)
(total emission);
{\it IR} data from from Knapp et al. (1989) and {\it X-ray} data 
from Fabbiano et al. (1992). The most recent estimate of the BH 
mass comes from Ferrarese, Ford \& Jaffe (1996), based on HST gas kinematics. 

{\bf NGC 4486B.} This is a low-luminosity companion of M87. The BH mass is
estimated by spectroscopic observations of Kormendy et al. (1996b), also
supported by the analysis of Lauer et al. (1996) of HST-WFPC2 images
suggesting that the nuclear emission could be fitted as an eccentric
disc (e.g. Tremaine, 1995). 

{\bf NGC 4374. } {\it Radio} data are from Dressler \& Condon (1978)
(total emission). {\it IR} data come from Condon et al. (1995), {\it
X-ray} data from Fabbiano et al. (1992). The value of BH mass is from
Bower (1996), using the HST velocity gradient across the central part of
the disc.

{\bf NGC 1316. } The {\it radio} flux is from the 1 Jy Catalogue 
(Kuhr et al. ), the {\it IR} one from Knapp et al. (1989), the
{\it X-ray} from Brinkmann et al. (1994). The value for the BH mass comes
from optical imaging of the central region of the galaxy by Shaya et al.
(1996) combined with limited kinematical information by Bosma et al. (1985). 
Flag 2 in last column is due to limited spatial resolution (1
arcsec) and poor kinematical data. 

{\bf NGC 7052. }{\it Radio} data are from GC91 and refer to the total
emission from the galaxy. The {\it IR} flux is from Knapp et al. (1989).
An upper limit to the value for the BH mass is reported by van den Bosch \& 
van der Marel (1995), based on gas kinematics
(turbulence with very high velocity, $v > 300$~km s$^{-1}$). They find that
the presence of a BH with such a mass is consistent with their data, 
but is not required. We keep NGC 7052 in our source list as providing a 
valuable limit on $M_{BH}$.

\bigskip\noindent
Altogether, 11 mass determinations for the galaxies in our sample are quite 
reliable (filled circles in the following figures), one uses gas kinematics
(open circle) and one surface photometry (filled triangle). 
\begin{table*}[htb]
\begin{center}
\begin{tabular}{lccccccccccc} 
\hline
  Object & d & $M_{B}$ & $F_{radio}^{tot}$ & $\log P_{5}^{tot}$ 
 & $F_{radio}^{core}$ & $\log P_{5}^{core}$ & $F_{60\mu}$ & 
  $F_{X}$ ($\times 10^{-13}$) & $\log L_{X}$ & $M_{\bullet}$ & QF \\ 
    & (Mpc) & & $(mJy)$ & (W/Hz) & $(mJy)$ & (W/Hz)&$(Jy)$&$(erg/cm^2/s)$& 
    $(L_{\odot})$ & $(M_{8})$ & \\
    1 & 2 & 3 & 4 & 5 & 6 & 7 & 8 & 9 & 10 & 11 & 12 \\ \hline
 {\bf M 31}     &0.7 &  -18.82 & 36 & 18.33  & 0.03 & 15.25 & 536.18 & 260.6  
& 5.60 &  0.3 & 3 \\
 {\bf NGC 3115} &10  &  -20.46 & $\le 0.33$ & $\le 18.60$ & $\le 0.33$ & $\le 18.60$ & $\le 0.13$ & $\le 6.41$ & $\le 
6.3$ & 20 & 3 \\
 {\bf M 32}     &0.7 &  -15.51 & $\le 7$ & $\le 17.61 $ & $\le 4$ & $\le 17.37$ & $\le 0.400$ & - & - & 0.03 & 3 \\
 {\bf NGC 4594} &16  &  -23.14 & 156 & 21.95 & 100 & 21.76 & 3.113  & 29.24 & 7.64 & 10 & 3 \\
 {\bf Milky Way}& -  &  -17.65 & 700 & 15.78 & 700   & 15.78 & -  & -  & - &  0.024 & 3 \\
 {\bf NGC 3377} &14  &  -19.74 & $\le 8.6$ & $\le 20.3 $ & $\le 0.5$ & $\le 19.06$ & 0.140 & $\le 1.36$ & $\le 5.91$ & 1.4 & 3 \\
 {\bf NGC 4258} &9   &  -20.47 & 170 & 21.21 & 1.8 & 19.24 & 17.340 & 45.66 & 7.06 &  0.7 & 3 \\
 {\bf M87}      &20  &  -21.86 & 59740 & 24.54 & 4000 & 23.37 & 0.3939 & 1050 & 
9.2  & 30 & 3 \\
 \\
 {\bf NGC 4261} &30  &  -21.87 & 4939  & 24.01 & - & - & 0.08  & 13.18 & 7.90 & 9.0 & 3 \\
 {\bf NGC 4374} &27  &  -21.42 & 2096  & 23 & 680 & 22.52 & 0.502  & 15.59 & 
7.29 &  $3.6 $ & 3 \\
 {\bf NGC 4486B}&20  &  -17.15 & - & - & - & - & - & - & - &  0.1 & 3 \\
 {\bf NGC 1316} &17  &  -21.76 & 49000 & 24.23 & - & - & 3.160  & 15.10 & 7.13 &  36 & 2 \\
 {\bf NGC 7052} &80  &  -21.22 & 122   & 23.01 & - & - & 0.460 & - & - & 
$\leq$9.0 & 1 \\
\hline
\end{tabular}
\end{center}
\caption{Table of the basic physical quantities for each galaxy. 
 {\bf Column 1. } Name of the object.
 {\bf Column 2. } Distance, based on models of the velocity field and
assuming H$_{0}=50$km\,s$^{-1}$\, Mpc$^{-1}$ if no distance indicators
are available. 
{\bf Column 3. } Absolute magnitude. m$_{B}$ values are taken from
{\sl NED} database.
 {\bf Column 4. }{\it Total} flux in radio band at 5\,GHz.
 {\bf Column 5. }{\it Total} radio power at 5\,GHz.
 {\bf Column 6. }{\it Core} flux in radio band at 5\,GHz.
 {\bf Column 7. }{\it Core} radio power at 5\,GHz.
 {\bf Column 8. } Far-IR flux at 60\,$\mu$.
 {\bf Column 9. } X-ray flux between 0.3 and 3.5 KeV.
 {\bf Column 10. } X-ray luminosity in solar units.
 {\bf Column 11. } Black hole mass in unit of $10^{8}$\, M$_{\odot}$.
 {\bf Column 12. } Quality flag, giving our estimate of the 
reliability of BH mass determination. 3: reliable estimate based on high
spatial resolution spectroscopy; 2: lower-resolution spectroscopy, 1: 
affected by turbulence.}
\end{table*}

\section{BASIC CORRELATIONS}

We analyse in this Section correlations between the central BH mass and
galaxy emissions at various wavelengths. These will provide useful
constraints when investigating the origin of the BH and the nature of the
gas flows in the nuclear region. Finally, such relations will be used for
statistical inferences about the occurrence and mass function of BHs.

A refined statistical approach is needed to deal with combined data and 
upper limits, and because of the very limited
statistics of our sample. Specifically, to evaluate the significance of the 
correlations, we use Cox's Proportional Hazard Model, the Generalized Kendall's
and the Spearman rank correlation statistics. Concerning the latter,
it is quite a robust correlation test but it is not very accurate for
limited samples like ours.
Altogether, we report in Figs. 1 to 4 the probability $P_{rand}$ to get by 
random the observed correlation from an uncorrelated parent population, as 
estimated with the generalized Kendall's $\tau$ test.. We have checked that 
in all cases the three different tests gave consistent results.
For all the details on these procedures we defer to Isobe, Feigelson, \& 
Nelson (1986).

Concerning the linear regression analysis, we used three different methods:
the classical Schmitt's regression analysis, the ``Expectation-Maximization''
algorithm assuming a normal distribution and the distribution provided
by the Kaplan-Meier estimator.
Values of the regression coefficients (also reported in the figures)
refer to the latter, as the best unbiased estimator (see Isobe, Feigelson, \& 
Nelson, 1986). 

We have to keep in mind that sometimes spurious correlations may be induced
by poorly controlled selection effects, in particular if the plotted 
quantities are distance-dependent (as it is our case). For this reason,
we will consider as more relevant for our analysis the remarkable 
differences among
the various correlation behaviours in Figs. 1 to 4, rather than the precise
values of the correlation significance and of the regression parameters 
for the separate plots.

KR95 found a good correlation between the black hole mass and the
absolute magnitude $M_{\rm B}$ of the bulge in the B-band. Bulge luminosities
coincide with the total ones for the new galaxies in our extended sample,
all being ellipticals. Figure 1 is the same plot of BH mass 
vs. $M_{\rm B}$ as in KR95, including the new galaxies in our sample. A
good correlation (0.4 per cent  probability of random occurrence, linear
correlation coefficient $CC \sim -0.84$) is confirmed, with a
residual scatter with respect to the formal regression line of
$\sigma_{mass}\simeq 0.54$.

\epsfxsize=8.5cm 
\begin{figure} 
\vspace*{-0pt} 
\hspace*{0pt}
\epsffile{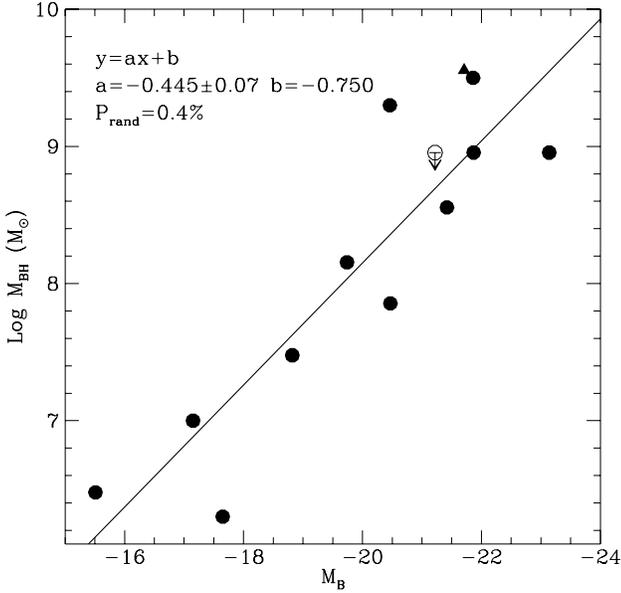} 
\vspace*{-0pt} \caption{Black hole mass
as a function of the absolute B magnitude of the bulge of the hosting 
galaxy. Filled circles,
triangles and open circles correspond to objects with QF=3, 2 and 1 in
Table 1. The coefficients of the linear fit and of the probability for
random occurrence from an uncorrelated parent distribution (see text) 
are indicated.}
\label{fig1}
\end{figure}

Remarkably different behaviours are displayed in Figure 2 for the 
plots of the BH mass versus the X-ray and far-IR luminosities. No
significant correlations are found in such cases, the emission being quite 
unrelated here to the presence of a nuclear supermassive BH.

\epsfxsize=8.5cm
\begin{figure}
\vspace*{-0pt}
\hspace*{0pt}
\epsffile{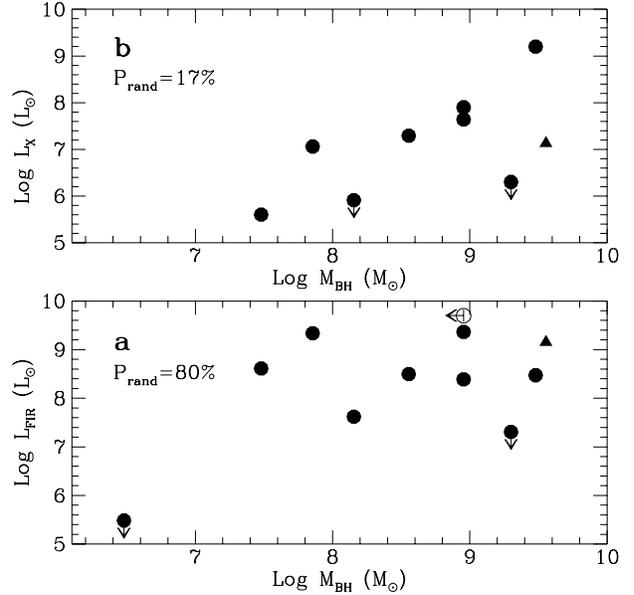} 
\vspace*{-0pt}
\caption{{\bf a)} Black hole mass versus the IR luminosity 
$\lambda L(\lambda) (60\mu m)$ at 60 $\mu$, in units of L$_{\odot}$. 
The symbols are as in Figure 1. {\bf b)} The same as a), 
for the X-ray luminosity.}
\label{fig2}
\end{figure}

For the radio flux, the most obvious relation with the BH mass is
expected from the high frequency compact nuclear emission. This is of
specific interest for the discussion about the ADAF model in the next
Section. Figure 3 is a plot of the 5 GHz core radio flux against BH mass
for the few galaxies for which high-resolution radio imaging is available
(only high quality, QF=3, mass values are used). In spite of the small
number of data points, there is a significant correlation (2.5 per cent 
probability of a randomly generated correlation) 
between core radio flux and BH mass.
Furthermore, quite a steep dependence is indicated by the data, the radio
power varying by 8 orders of magnitude for a variation of the BH mass of
$\sim 3$ orders of magnitude. In particular, there is a $3\sigma$
difference with the almost linear correlation found in Fig. 1.

Finally, a remarkably tight correlation (0.25 per cent probability of random
occurrence excluding NGC 3115, see below, 0.7 per cent including it) is
revealed between BH mass and the 5 GHz total radio emission. From the radio
flux we have subtracted here the contribution due to star-formation
activity, estimated from the well-established relationship between far-IR
and radio emission (e.g. Walsh et al. 1989; Helou et al. 1985):
\begin{equation}
 L_{\rm 5 GHz} = 0.01\ L_{\rm 60 \mu m}.
\end{equation}%
This has allowed us to derive the radio emission of non-thermal origin
arising from the nuclear activity. Note that the contribution from
star-formation is significant for only the later-type galaxy M31 and NGC 4258
(NGC 3115 has only upper limits and nothing can be said). 
In this case too, a steep dependence of the radio
flux on the BH mass is displayed ($P_5 \propto M_{\rm BH}^{2.66}$), very
similar to the one inferred for the core radio fluxes.

\epsfxsize=8.5cm
\begin{figure}
\vspace*{-0pt}
\hspace*{0pt}
\epsffile{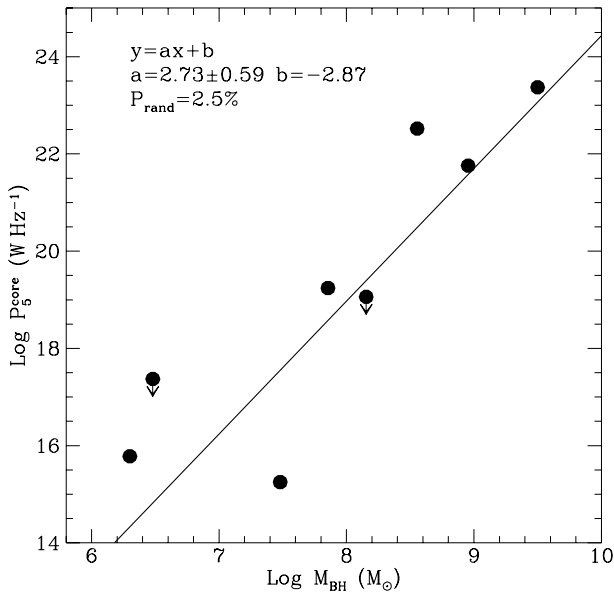} 
\vspace*{-0pt}
\caption{Black hole mass as a function of the core radio power at 5 GHz, 
in units of [Watt/Hz], for the few objects with reliable BH mass
determinations (QF=3) and for which radio imaging photometry is
available. The results of the regression analysis appearing in the figure
are based on a survival analysis technique.} 
\end{figure}

The most significant deviation from the best-fit lines in Fig. 4 concerns
NGC 3115. Its inferred BH mass is very similar to the one in M87, but the
radio flux is six orders of magnitude lower, which is likely due to an
almost zero mass accretion rate in NGC 3115. This seems also to be
confirmed by the very small amounts of halo gas mass in this galaxy, as
inferred from the low X-ray luminosity
(L$_{X}\sim1.35\times10^{40}$\,erg\,s$^{-1}$). NGC 3115 should then be in
a very inactive stage, with a starving black hole left over from a much
brighter past AGN phase.
The significantly deviant behaviour of NGC 3115 in the plot of Fig. 4
has convinced us to exclude it from regression analyses (see parameters
reported in Fig. 3 and 4) aimed at testing the ADAF model and estimating
the remnant BH mass function. This is justified in both cases: in the former,
NGC 3115, without appreciable accretion, cannot test the ADAF; in the latter,
our mass function will not account for the small fraction of BHs without
even low-level radio activity (see Discussion).

In principle, we would expect the high-frequency core radio flux to be at
least as good a tracer of the BH mass as the total radio emission, also
in the light of the current interpretation for such a nuclear radio
activity, see Sect. 4 below. In practice, we see in Fig. 3 a larger
scatter than in Fig. 4, with an otherwise similar overall dependence. The
scatter in Fig.3 is likely to be attributed to the sparser nature of the data,
and to the fact that core radio fluxes are highly variable, by factors up
to 10, on time scales as short as a week. Further high
resolution, simultaneous, multi-wavelength radio imaging for nearby 
early-type galaxies is obviously needed.

\epsfxsize=8.5cm
\begin{figure}
\vspace*{-0pt}
\hspace*{0pt}
\epsffile{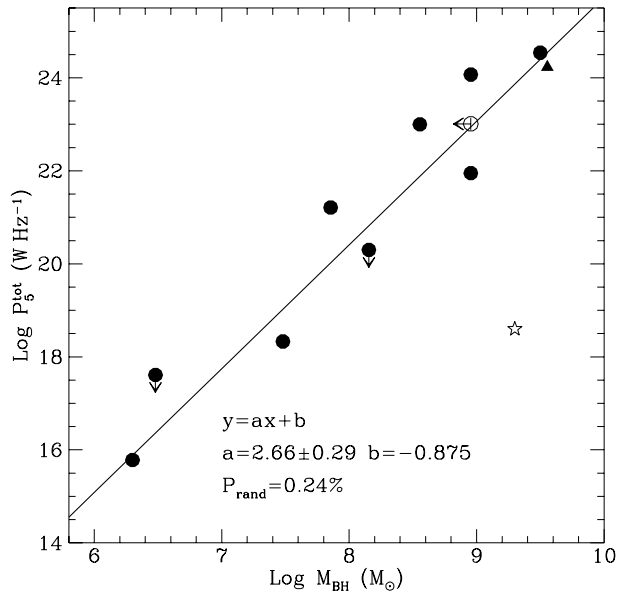} 
\vspace*{-0pt}
\caption{Black hole mass as a function of the total radio luminosity at 
5 GHz, in units of [Watt/Hz]. The star at the lower right corresponds to
NGC 3115.  
}
\label{fig3}
\end{figure}

\section{A TEST FOR THE ADVECTION-DOMINATED ACCRETION FLOW MODEL} 

The results of the regression analyses in the previous Section may have
an interesting impact on a recently proposed accretion mode based on the
concept of advection-dominated accretion flows (see Narayan \& Yi 1995;
Abramowicz et al 1995; Narayan \& Yi 1996). In this model, the ions are
close to the virial temperature, $T_{i}\sim10^{12}r^{-1}$ K, where $r$ is
expressed in Schwarzschild radii, while the electrons are much cooler,
although trans-relativistic, $T_{e}\sim10^{9}-10^{10}$ K. This occurs at
low accretion rates ($\dot{m}\le0.3\alpha ^2$, where
$\dot{m}=\dot{M}/\dot{M_{Edd}}$ is expressed in Eddington units, and
$\alpha$ is the standard viscosity parameter, Shakura \& Sunyaev 1973),
when Coulomb collisions are unable to transfer energy from ions to the
electrons as fast as it is released by viscous forces. Most of the energy
of the ions is transported into the black hole.

The process can be described in terms of the various relevant timescales
in the accreting plasma: $t_{\rm visc}$, the viscous timescale accounting
for heating of the ions, $t_{\rm p-e}$ the Coulomb collision timescale
cooling timescale of ions and heating of the electrons, and $t_{\rm
cool}$, which is the cooling timescale for the electrons. Under the low
accretion rate hypothesis, we have:
\begin{equation}
  t_{\rm cool} \ll t_{\rm visc} \ll t_{\rm p-e}\,,
\end{equation}
which means that most of the energy remains with the ions. The
electrons radiate efficiently by bremsstrahlung and synchrotron emission,
but because of the weak Coulomb interaction, this radiation cannot drain
all the energy from the ions.

Since the ions are close to the virial temperature the flow is spatially
thick and resembles a torus.  Significant radio emission is predicted
from the torus due to cyclo-synchrotron radiation from hot electrons in
the (expected) equipartition magnetic field. Fabian \& Rees (1995) use
this to explain part of the radio emission from early-type galaxies
(Sadler et al. 1989; Wrobel \& Heschen 1991). Other applications of this
model can be found in Narayan, Yi \& Mahadevan 1995 (Sagittarius
A$^{\ast}$), Lasota et al. 1996 (NGC 4258), Reynolds et al. 1996 (M87),
Di Matteo \& Fabian 1996b (NGC 4649).

There are some very important features in the ADAF model. a) If the
energy is to be advected with the plasma without being released, it is
necessary that {\it the central object has no hard surface}, so it has to
be a BH. b) This model requires a {\it non-Newtonian gravitational field}
close to the central mass, as expected from general relativity. 
c) The geometry of the torus can favour, near the rotation axis, the
formation of 
jet-like structures (Fabian \& Rees, 1995).

The ADAF accretion mode would then allow, if confirmed, quite 
fundamental tests of the BH paradigm. Therefore it would be of key  
importance to prove -- or disprove -- it.

A striking difference between the plots in Figs. 1 and 3 is in the slope
of the best-fit regression lines. The fit with the optical data ($M_{\rm
BH}-M_{\rm B}$) implies an almost linear dependence of the bulge optical
luminosity on the mass of the dark object
\begin{equation}
L_{\rm B} \propto M_{\rm BH}^{4/5} . 
\end{equation}
This may indicate a simple scaling of the mass of the putative central BH
on that of the host galaxy, something clearly related with the 
formation process.

In contrast, much steeper dependences of the high-frequency radio 
luminosity on the mass of the dark object are seen in Figs. 3 and 
4 comparing $M_{\rm BH}$ with both the total and the core radio fluxes:
\begin{equation}
L_{\rm 5 GHz} \propto M_{\rm BH}^{2.2-3.0}.
\end{equation}
This shows that the radio flux is not only a good tracer of the presence
of nuclear massive dark objects in otherwise inactive galaxy nuclei, but
also a very sensitive indicator of the BH mass.

Concerning the physical origin for the very steep dependence of the radio 
flux on $M_{\rm BH}$, it cannot be simply explained 
in terms of a generic scaling of BH mass on that of the host galaxy,
as in the case for the optical regression.
It rather requires a specific process originating it. It can be 
accounted for in the ADAF scheme, as shown below.

In the ADAF, $\dot M \propto R^2 n v$, where the velocity $v\propto M_{\rm
BH}^{1/2}/R^{1/2}$ and the density is $n$ (from continuity and the inflow
velocity which is proportional to the orbital velocity). The ion temperature
is close to virial $T_i\propto M/R$ so an equipartition magnetic field
$B\propto R^{-5/4}\dot M^{1/2} M_{\rm BH}^{1/4}$. The high frequency cutoff
of the self-absorbed cyclotron emission occurs at an approximately fixed
factor above the cyclotron frequency (Narayan \& Yi 1995), since this is
where the emission locally peaks $\nu\propto B$. Thus we can write $R\propto
\dot M^{2/5} M_{\rm BH}^{1/5}/\nu^{4/5}$. Now the radio power $L\propto
\nu^2 R^2$ (Rayleigh-Jeans emission) therefore $L\propto \nu^{2/5}\dot M^{4/5}
M_{\rm BH}^{2/5}$ (Mahadevan 1997).

In the ADAF model for elliptical galaxies, it is assumed that the hot
interstellar medium accreting onto the central black hole has little angular
momentum and thus that the Bondi accretion solution applies. The accretion
rate is then determined by the mass of the black hole, the density of the
hot surrounding gas, $n$, and its temperature (or equivalently its sounds
speed, $c_{\rm s}$); $\dot M \propto M_{\rm BH}^2 n/c_{\rm s}^3$. This means
that $L\propto
\nu^{2/5} M_{\rm BH}^2 n^{4/5} c_{\rm s}^{-12/5}$.

Assuming that $n$ is controlled by cooling and stellar mass loss it is
plausible that $n \propto M_{\rm gal}$. Then noting from eq.(3) that
$M_{\rm BH} \propto M_{\rm gal}$ and using the Faber \& Jackson (1976)
relation, we infer that $c_{\rm s}
\propto M_{\rm BH}^{1/4}$ so the final result is
\begin{equation}
L_{\rm radio} \propto \nu^{2/5} M_{\rm BH}^{2.2} .
\end{equation}
This is in reasonable agreement with the observed dependence on black hole
mass. If there is little gas in the galaxy, perhaps due to an energetic
event which created a wind, then the relation $n \propto M_{\rm gal}$
assumed above is broken and little radio emission is expected. NGC 3115 is
therefore not inconsistent with our model, despite the apparent lack of
agreement in the correlations presented in Section 3. 

We note that any self absorbed process in which the emission was mostly
dependent on the emitting area available around a black hole would show
$L\propto M_{\rm BH}^2$. It may therefore be possible to generate a jet
model accounting for the correlation in Fig. 4. Much more knowledge
would be required, however, on the density and power of a jet as a
function of galaxy and accretion parameters than is currently available. 

\section{THE BLACK HOLE MASS FUNCTION}  

The very tight correlation reported in Fig. 4 shows that the total radio
power is a very sensitive function of the mass of the nuclear black hole.
Given the tight relationship of BH mass and total radio flux and the fact
that rich statistics and complete surveys exist for the total
centimetric radio emission from galaxies, we use in the following the
radio regression of Fig. 4 to derive the BH mass function in the local
universe.

A formal solution for the latter would be first to compute the bivariate
luminosity distribution, i.e. the conditional probability distribution
to have a value for the radio luminosity $L_{\rm 5 GHz}$ within a given
interval for any given value of the BH mass $M_{\rm BH}$. Then the BH mass
function would be given by the convolution of the bivariate luminosity
distribution with the radio luminosity function (LF).
This full procedure, however, is not allowed by the very limited statistics
available to us on BH masses.

On the other hand, the tightness of the empirical relation 
$\log P_5^{\rm tot}-M_{\rm BH}$ 
implies that a simple transformation of the radio LF via eq. (3)
provides a good account of the mass function: 
 \begin{equation} 
\frac{d\,N(M_{\rm BH})}{d\,\log(M_{\rm
 BH})}=\frac{d\,N(P_{5})}{d\,\log(P_{5})} \times
 \frac{d\,log(P_{5})}{d\,\log(M_{\rm BH})}\,, 
\end{equation}
where the first term in the right end side is the radio luminosity function, 
and the second one is the slope of the best-fit regression in Fig. 4.
We exploited the regression with the total radio power of Fig. 4, instead of
the one with core radio flux, because only for the former well defined and
reliable luminosity functions do exist.

As for the latter, we used the one at centimetric wavelengths reported by
Franceschini et al. (1988) and Toffolatti et al. (1987).
While this reference radio LF is defined at 2.4 GHz, most of the radio data 
in Fig. 4 are defined at 5 GHz. 
For the spectral conversion, we used the scaling
S$_{\nu}$ $\sim \nu^{-0.75}$, which applies on average to total radio power
(usually dominated by extended emission) from radio galaxies.

As a check, we have independently estimated the BH mass function from
the ($M_{\rm B}-M_{\rm BH}$) relation (see eq.[3]) and using the B-band optical
luminosity function of field ellipticals and S0's reported by
Franceschini et al. (1988), following the same procedure as for the
radio case. 

Figure 5 shows the BH mass functions derived with the two procedures.
They are fairly consistent over the bulk of the distribution, with
significant deviations at low and high values of the mass $M_{\rm BH}$. 
Given the much better defined relation of $M_{\rm BH}$ with the radio
flux and the more precise physical interpretation, the BH mass
function based on the radio correlation is to be taken as our best
guess.

 \epsfxsize=8.5cm
 \begin{figure}
 \vspace*{-0pt}
 \hspace*{0pt}
 \epsffile{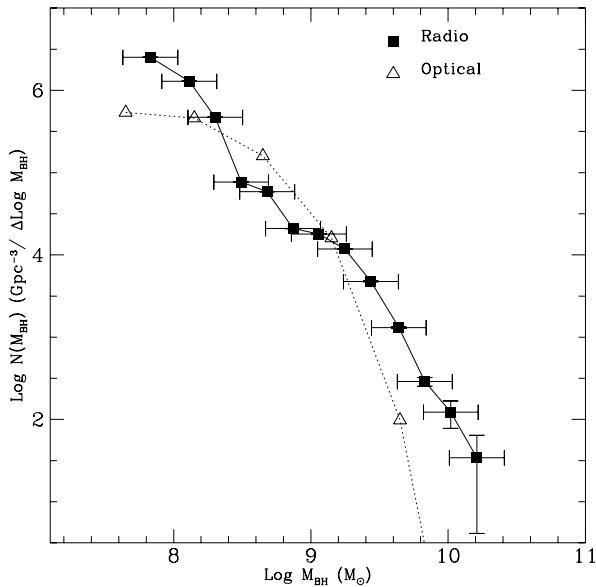} 
 \vspace*{-0pt}
 \caption{Black hole mass functions derived as explained in text. 
Solid squares refer to the evaluation based on the radio-$M_{BH}$ 
correlation, while open triangles to the optical relation.
Vertical errorbars are purely statistical uncertainties coming from the 
radio luminosity function, while horizontal bars reflect the uncertainty
in BH mass determination as shown by the residual distribution in Fig. 4.
}
 \label{fig5}
 \end{figure}

Integrating the mass function based on the radio regression in the 
$M_{BH}$ interval over which it is defined, we estimate a total mass 
density of massive black holes in galactic nuclei 
\begin{equation}
\rho_{\rm BH}\simeq 1.2\times10^{5}$\,M$_{\odot}$\,Mpc$^{-3}.
\end{equation}
This value is in good agreement with the Chokshi \& Turner (1992) 
prediction based on quasar optical counts.  

A stronger requirement on BH-remnant mass density ($\rho_{\rm BH}\sim
4\times10^{5}$\,M$_{\odot}$\,Mpc$^{-3}$) has been estimated by Granato
et al. (1997) in order to explain the hard X-ray background (HXRB) as the
integrated contribution of evolving self-absorbed Active Galactic Nuclei.
Our result in eq. (7) independently confirms the old argument that
luminous quasars contribute only a
small fraction of the XRB: in our version of the argument, super-massive
BHs with $M_{\rm BH} > 10^8\ M_{\odot}$ can explain only roughly 20-30
per cent of the XRB.

However, given the steep dependence of $N(M)$ in Fig. 5:
\begin{equation}
N(M_{\rm BH}) \propto M_{\rm BH}^{-2}, 
\end{equation}
the total mass density $\rho_{\rm BH}(>M_{\rm BH}) \propto M_{\rm
BH}^{-1}$ is weighted by low mass values. A slight extrapolation of
$N(M_{BH})$ below the limit $M_{BH}=10^8\ M_{\odot}$ can easily fill in
the gap to the value required to explain the HXRB energetics.  This
supports the idea that the emission from a numerous (several $10^{-3}$
objects Mpc$^{-3}$) population of low-luminosity AGNs powered by
moderately massive BH ($M_{\rm BH} \sim 10^7\ M_{\odot}$) has produced
the HXRB. As shown by the morphological type of the host galaxies of BH in
the lower left of Figs. 3 and 4, and as implied by the required space
densities, a sizeable fraction of these AGN should be hosted in
early-type spirals. After Fig. 1, their BH mass should be roughly
proportional to the mass and luminosity of the bulge component.

\section{DISCUSSION }

Although the dataset on nuclear super-massive black holes used in the
present analysis is still quite sparse and incomplete, some interesting
trends can already be observed in the available data. More information is
to be expected soon, in particular from the refurbished Hubble Space
Telescope and from high spatial resolution ground-based observations.

If we take the present sample as representative of the normal galaxy
population (specifically early-type galaxies), then some interesting
consequencies may be inferred.

A good correlation is confirmed between the black hole mass and the
absolute magnitude, and thus baryon mass, of the spheroidal component of
galaxies, as found by KR95. A roughly linear dependence is indicated for
the BH mass on the mass of the host spheroid, which may provide crucial
information relating to the origin of the BH with that of the host
galaxy. A strong constraint may be implied for the nuclear gas dynamical
processes occurring in the primeval forming spheroid or in the early
merger events producing it.

In spite of the small number of objects for which data are available and
of the noise introduced by the time variability of the high-frequency 
core radio power, the latter shows a good correlation with and a steep 
dependence on $M_{\rm BH}$. 
If compared with the remarkable absence of any dependences of
$M_{\rm BH}$ on the IR and X-ray luminosities (Figs. 2), this seems to
support the concept of an advection-dominated black hole accretion in the
core of these galaxies. Indeed, such model predicts strongly peaked
emissions in the high-frequency radio and hard X-ray, and essentially no
contribution to the far-IR, optical and soft X. Furthermore, the observed
steep dependence of the core radio luminosity on $M_{BH}$ (see Fig. 3 and
eq. [4]) is well reproduced by the ADAF model.

The total radio flux shows the best correlation with the mass of the
central dark object (see  Fig. 4). 
The scatter is low over a wide range of power values.
Only one datapoint, that corresponding to NGC 3115, falls clearly
outside the relation in Fig. 4. Even such a low accretion-rate 
needed to sustain an advection-dominated disk is ruled out by this
observation. The very massive BH in NGC 3115 is not accreting gas at all.
It is beyond the scope of this paper to estimate how many of these 
starving BHs could be missed even by a sensitive radio survey.
Certainly other similarly deviating data points should exist.
The only thing we can say now is that they should be a minority population
among early-type galaxies, if our poor statistics makes sense after all
(one deviating object out of 12).

The relation in Fig. 4, and potentially the one in Fig. 3 when better 
data will be available, provide more powerful tools than
the optical one to investigate the nature of the Massive Dark Objects
and for statistical inferences about the BH occurrence and mass 
function. Using it, we have calculated what we believe to be a fairly 
reliable black hole mass function
and the total mass density of remnants in normal galaxy nuclei. 
It turns out to be basically consistent with predictions of models 
interpreting the quasar and AGN activity as short-lived emission events,
as opposed to the classical view of a continuity between high-redshift 
quasars and low-redshift AGNs, which would imply a prolonged activity
over a several Gyr timescale and would violate the constraints by the
observed M/L ratios in AGNs.

We report in Figure 6 a comparison between our estimated remnant BH
mass function and predictions by Cavaliere \& Padovani (1988, 1989).
It is evident that curve C, corresponding to a continuous long-lived
activity pattern, is very significantly excluded by the data.
The latter are instead remarkably consistent with the case of a short-lived
or recurrent activity.

 \epsfxsize=8.5cm
 \begin{figure}
 \vspace*{-0pt}
 \hspace*{0pt}
 \epsffile{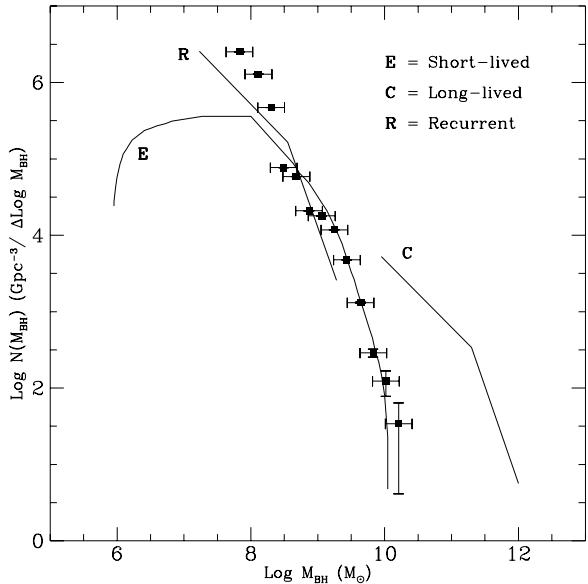} 
 \vspace*{-0pt}
 \caption{
Comparison of our estimated remnant BH mass function with the predictions of
various patterns for AGN activity, as discussed in Cavaliere \&
Padovani (1988, 1989). 
}
 \label{fig6}
 \end{figure}

Our estimated BH mass function has also a relevant impact on related
questions such as the origin of the HXRB. Assuming it to be contributed by
self-absorbed AGNs, then a numerous population of moderately massive BH's
should exist with $M_{BH} \sim 10^7\ M_{\odot}$. This implies that a good
fraction of early-type spirals (those in which the spheroidal component is
more prominent) harbors a BH of roughly this mass. Those few achieving a
sustained rate of gas accretion (e.g. due to a merger or a close
interaction) are visible as Seyfert nuclei, while the large majority
accreting at low rates are either emitting in the radio band via an
advection-dominated disc, or entirely inactive. In any case, low-mass
low-luminosity AGNs are likely to origin the HXRB.

Many observational requirements are indicated by our analysis. Concerning
the "demography" of super-massive BHs in local galaxies, one could
operate as follows. The first step would be to perform high resolution
simultaneous multi-frequency radio mapping of a sample of nearby 
(mostly early-type)
galaxies. This is easy to achieve e.g. with the VLA. Then, following Fig.
3, we can select objects showing enhanced core radio emission for high
spatial resolution spectroscopic observations. 

Concerning the other issue discussed here, i.e. proving the ADAF model,
a testable feature is the weakly inverted radio-mm spectral shape 
($f_\nu \propto \nu ^{0.4}$). Again high resolution, high frequency
radio observations, coupled with millimeter and sub-millimeter
observations (now relatively easy with the very sensitive existing mm
telescopes) of a sample of galaxies selected as mentioned above, may
soon provide a very promising way to test it. 

Finally we note that very deep radio surveys (e.g. Gruppioni et al 1996;
Hammer et al 1995) find that a significant fraction of sources below 1
mJy are associated with elliptical galaxies at redshifts of about 1.
These may be more distant examples of what we have studied in this paper
and offer the possibility to explore the evolution of the mass and
distribution of black holes with cosmic epoch.

\section*{Acknowledgments}
We have benefited by discussions with A. Cavaliere, L. Danese and T. Di
Matteo. We are indebted to the referee, Frank van den Bosch, for significant 
help in improving the paper, and for pointing out the impact of time 
variability in the high-frequency core radio flux.


\end{document}